# Square Maxwell's fish-eye lens for broadband achromatic super-resolution imaging

Jue Li(李珏)[1,2] ,Yangyang Zhou(周杨阳)[1,2,*] ,and Huanyang Chen（陈焕阳）[1,2,**]

[1]Institute of Electromagnetics and Acoustics and Department of Physics, Xiamen University, Xiamen 361005, China
[2]Fujian Engineering Research Center for EDA, Fujian Provincial Key Laboratory of Electromagnetic Wave Science and Detection Technology, Xiamen Key Laboratory of Multiphysics Electronic Information, Xiamen University, Xiamen 361005, China

* corresponding anthor: zhouyangyang@stu.xmu.edu.cn
**corresponding anthor: kenyon@xmu.edu.cn

**Abstract:** Broadband super-resolution imaging is important in optical field. To achieve super-resolution imaging, various lenses from superlens to solid immersion lens have been designed and fabricated in recent years. However, the imaging is unsatisfactory due to low work efficiency and narrow band. In this work, we propose a solid immersion square Maxwell's fish-eye lens which realizes broadband (7-16GHz) achromatic super-resolution imaging with full width at half maximum around 0.2λ based on transformation optics at microwave frequencies. In addition, a super-resolution information transmission channel is also designed to realize long-distance multi-source super-resolution information transmission based on the super-resolution lens. With the development of 3D printing technology, the solid immersion Maxwell's fish-eye lens is expected to be fabricated in high frequency band.

## 1.Introduction

Microscopy is a significant tool for research about life science and natural science. The resolution of conventional lens system is constrained above one-half wavelength, due to the diffraction limit that evanescent waves carrying tiny information of an object decay in the far-filed. Many efforts have been made to overcome the diffraction limit, one important step is the proposed perfect lens in 2000 [1]. The perfect lens can achieve perfect imaging by negative refractive materials to restore amplitude of all evanescent waves and phase of all propagating waves at the imaging plane. Following this idea, a series of superlenses [2-4] was designed from microwave to optical band [5, 6]. Later on, hyperlens as far-field magnified super-resolution lens was proposed by hyperbolic dispersion materials, which can transform evanescent waves into far-field propagating waves. Utilizing alternative metal and dielectric materials, one-dimensional hyperlens[7], two-dimensional hyperlens [8] have been designed and fabricated. However, for superlens and hyperlens, it is difficult to push forward to applications and relative experimental works were rarely reported in recent years due to the challenge of fabrication and the big loss from plasmonic resonance mechanism and impedance mismatching between the object and lens.

On the other hand, utilizing total internal reflection (TIR) happening at the interface to excite evanescent waves, solid immersion lens (SIL) [9, 10] were suggested to overcome the diffraction limit for super-resolution imaging. They have been studied widely through the application of high refractive-index (RI) solid material and specific geometric optical design [11-13]. With the development of materials sciences, different types of SILs were design and fabricated from conventional structures to novel metamaterials structures [14-16]. However, a major drawback of SILs is their narrow band imaging capability due to aberration.

Recently, absolute instruments with a gradient RI (GRIN) profile [17-24] have been developed rapidly for their excellent capability to control the propagation of electromagnetic waves [25-30] and especially geometrical perfect imaging. Among these absolute instruments, Maxwell's fish-eye lens (MFEL) [31] has attracted a great of attention due to its property of geometrical perfect imaging optics and big numerical aperture (NA). Many applications, based on the MFELs, were designed and fabricated from microwave frequency to optical frequency [32]. Significantly, combining the solid immersion mechanism with geometrical perfect imaging of MFEL, a solid immersion MFEL as a near-field super-resolution lens [33, 34] has been designed and fabricated for super-resolution imaging at microwave frequencies. However, due to imaging along the lens's curved surface, there is imaging distortion which limits the structure shape of the imaging object for further applications.

In this work, we utilize a power tool transformation optics [35-37] to design a solid immersion square MFEL, which realizes super-resolution imaging and overcomes the disadvantage of the solid immersion circular MFEL. This paves the way to further application about solid immersion MFEL. Utilizing a Schwartz-Christoffel (SC) mapping [38], the solid immersion circular MFEL is transformed into a solid immersion square MFEL to achieve a wide-band (7-16GHz) high-resolution imaging effect with FWHM around 0.2λ. By commercial software COMSOL multiphysics, we numerically calculate the functionalities of the lens and prove the excellent imaging ability of high resolution and wideband for the proposed solid immersion square MFEL.

## 2. Square MFEL for high-resolution

Let us recall a MFEL with a GRIN distribution that can realize perfect imaging. Assuming a MFEL located in virtual space ($w$ space), it RI distribution can be expressed as follows [39]:

$$n_w = \frac{2n_0}{1+\left(\frac{r_w}{R}\right)^2} \quad (1)$$

where $r_w$ is the distance from the center of the lens. $R$ is the radius of the lens, $2n_0$ denotes the RI at the center of MFEL, and $n_0$ represents the ambient RI. Fig. 1(c) shows the RI distribution of MFEL where the lens radius is 1(the unit is arbitrary), and $n_0$ is set to 1. The MFEL can realize aberration-free imaging from point to point in the lens [22]. It also can achieve super-resolution imaging in wave optics via introducing a solid immersion mechanism [34]. Although super-resolution imaging can be achieved through solid immersion MFEL, its application scope is limited due to the curved boundary of the lens, leading to distortion of the imaging. To extend the application prospects of the solid immersion MFELs, the curved edge of the MFEL is considered to be changed into a straight edge according to the theory of conformal mapping. Such design of a solid immersion square MFEL is obtained based on the solid immersion circular MFEL by utilizing a SC mapping [38] as follows:

$$q(z) = 1 - \frac{1}{\left(sn\left(\frac{2i\sqrt{2}\pi^{3/2}\left[z+(1-i)\right]}{\Gamma\left(-\frac{1}{2}\right)^2}\bigg|2\right)\right)^2} \quad (2)$$

$$\omega(q) = \frac{q(z)-i}{q(z)+i} \quad (3)$$

where $sn(\varphi|m)$ represents the Jacobian elliptic function with the modulus of the ellipse $m$ and $\Gamma$ is the gamma function.

According to Eq. (2), we choose a transformation from the upper half-plane (coordinates $q$) to the square in physical space (coordinates $z$). Another conformal mapping, which is from the unit disk (coordinates $w$) to the upper half-plane (coordinates $q$) can be implemented based on Eq. (3). Consequently, a SC transformation from the unit disk to the square can be accomplished by taking the two equations together, which is the conformal mapping from a circular MFEL onto the square MFEL. The transformation relationship can be easily seen in Figs. 1(a)-(b), where the orthogonal mesh in the constant radial and azimuthal coordinates ($r-\theta$ directions) within a unit disk in Fig. 1(a) is transformed into the orthogonal mesh inside the square in Fig. 1(b) by the SC mapping. Thus, the

transformed RI distribution in physical space can be obtained as follows:

$$n_z = n_w \left|\frac{dw}{dz}\right| \qquad (4)$$

Here, the RI distribution of the square MEFL is calculated by Eq. (4) based on a circular MFEL with $n_0 = 1$, as shown in Fig. 1(d). According to the theory of TO, it can be inferred that the solid immersion square MFEL can also achieve super-resolution imaging which is similar to the solid immersion circular MFEL.

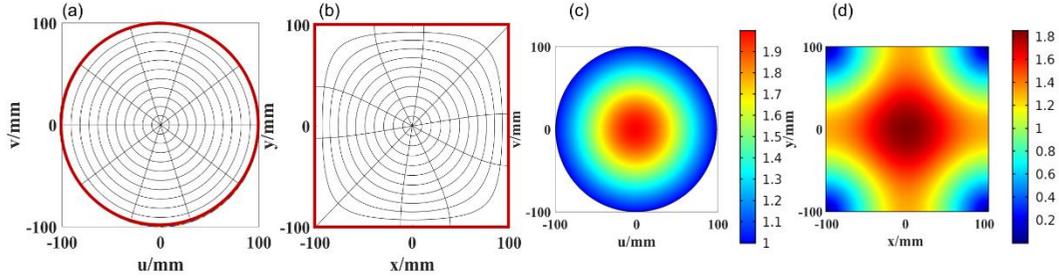

**Fig. 1.** (a) Coordination mapping in virtual space ($w$ coordinate). (b) Coordination mapping in physical space ($z$ coordinate). (c) The RI distribution of MFEL in virtual space ($w$ coordinate). (d) The RI distribution of the square MFEL according to SC mapping in physical space ($z$ coordinate).

To analysis the influence of the TIR mechanism for super-resolution effect in the solid immersion square MFEL, the value $n_0$ is artificially changed from 1 to 3.1 at an interval of 0.7 to explore the relationship between the value of $n_0$ and the imaging effect of the square MFEL as shown in Figs. 2(a)-(d). The focusing effect of the square MFEL with $n_0$ of 1, 1.7, 2.4, and 3.1 at the frequency of 15GHz is simulated by method of full-wave numerical simulation respectively. A point source (or a line current source for transverse electric polarization of light) is placed at $x = -100$ mm, $y = 0$ mm (the center of the square MFEL is at the origin) as an extinction source. In the figure, there is a sub-wavelength light spot at the corresponding point of the square MFEL respectively. The white dotted line represents the observation position of the imaging point ($x = 101$ mm, $y = -100 \sim 100$ mm), which is not at the boundary of the square MFEL but in the air near the boundary of the lens. The red curve is the normalized electric field intensity of the imaging point, and the corresponding FWHM is calculated. It can be observed from Figs. 2(a)-(d) that with the increasing of the value of parameter $n_0$, the TIR phenomenon at the interface between the square MFEL and the air is gradually enhanced. Therefore, the evanescent wave is excited by the TIR mechanism at the interface of lens/air, and the values of the corresponding FWHM at the imaging point gradually decrease from 0.34 $\lambda$ to 0.16 $\lambda$, which is much less than the diffraction limit. This manifests that the solid immersed square MFEL also has excellent super-resolution imaging capabilities and can achieve the same effect as the solid immersion circular MFEL. The relative electric field patterns are shown in Figs. 2(e)-(h) which also well explain the good imaging performance of the square MFEL due to the almost completely axisymmetric field distribution. Consequently, through the above analysis, it is noted that with the increase of $n_0$, the super-resolution imaging effect of the square MFEL gradually improved by the TIR mechanism.

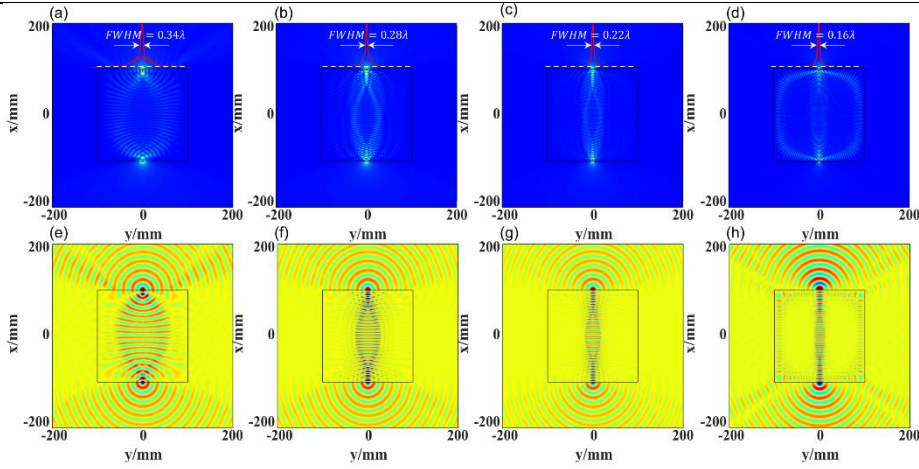

**Fig. 2.** Imaging performance of the square MFEL with different $n_0$. (a)-(d) The related electric field intensity distribution and the corresponding FWHM of the square MFEL at 15GHz with $n_0$ of 1, 1.7, 2.4 and 3.1, respectively. (e)-(h) The relative electric field $E_z$ distributions respectively.

Next, we will discuss the frequency response of the super-resolution effect in the square MFEL. A square MFEL with $n_0 = 2.4$ is selected for analyzing the broadband imaging performance of the solid immersion square MFEL at different frequencies from 7GHz to 16GHz. A point source is placed at $x = -100$ mm and $y = 0$ mm to excite cylindrical transverse electric (TE) wave. The imaging effects of the solid immersion square MEFLs at different frequencies are observed at $x = 101$ mm and $y = -100 \sim 100$ mm (located at the air background). Figs. 3(a)-(d) illustrate the electric field intensity $|E^2|$ and the corresponding FWHM of the solid immersion square MFELs at frequencies of 7GHz, 10GHz, 13GHz, and 16GHz, respectively. It can be seen that the corresponding FWHMs at frequencies of 7GHz, 10GHz, 13GHz, and 16GHz are all less than 0.25λ, which is far below the diffraction limit. From the figures, it is noted that part of super-resolution imaging is located at the air background and a slight resonance emerges in the lens compressing the resolution. At the same time, Figs. 3(e)-(h) show the corresponding electric field distribution patterns. From the above analysis, it is found that the solid immersion square MFEL can achieve super-resolution imaging at most frequencies. However, the super-resolution imaging is invalid at other specific frequencies due to the destruction from stable whispering gallery mode (WGM) [34], which is a resonance mode due to continuous TIR of the impedance mismatching at the interface of lens/air. To circumvent the disadvantage of WGM, an optimal design, transforming a circular MFEL into a quasi-square MFEL utilizing the SC mapping, is proposed as shown in Figs. S1(d)-(f) (more details see the Supplement 1 for supporting content). From the figures, it is seen that WGM is eliminated by transforming the circular MFEL into quasi-square MFEL. The simulated results show that the solid immersion quasi-square MFEL can achieve better super-resolution imaging than that of solid immersion square MFEL at frequencies of 8GHz, 10GHz, and 12GHz, respectively.

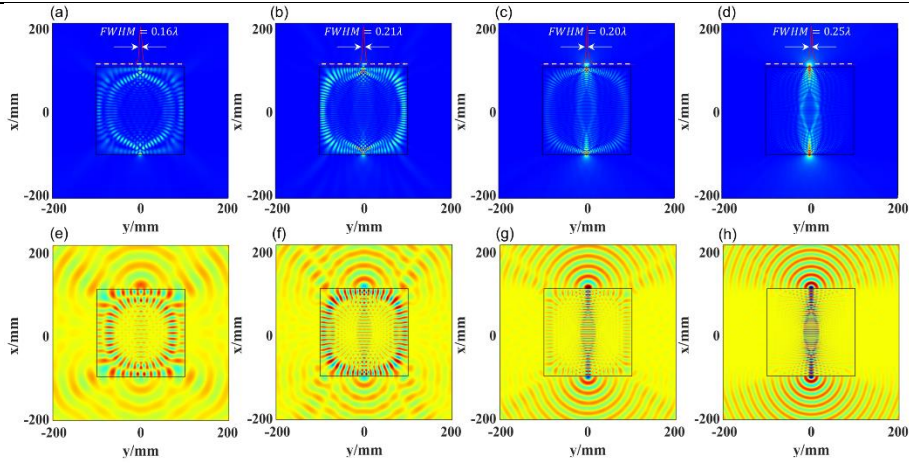

**Fig. 3.** Broadband imaging effect of solid immersion square MFEL at different frequencies. (a)-(d) The electric field intensity and the corresponding FWHM of the solid immersion square MFEL at frequencies of 7GHz, 10GHz, 13GHz, and 16GHz, respectively. (e)-(h) The corresponding real part of electric field distribution respectively.

Furthermore, to better illustrate the super-resolution effect of solid immersion MFELs, the imaging performances of a solid immersion square MFEL with $n_0 = 2.4$ and a square conventional MFEL with $n_0 = 1$ are compared with two exciting point sources placed with a distance of spacing $\lambda/3$ at the frequencies of 8GHz, 9GHz, 11GHz, and 12GHz respectively, as shown in Fig. 4. The relative distribution curve of electric field intensity in the air is calculated, and the imaging quality of square MFELs for two different $n_0$ is analyzed. Figs. 4(e)-(h) reveal that the square conventional MFELs with $n_0 = 1$ fail to resolve the two points spacing less than a one-half wavelength and only one focal point is presented from the electric field intensity curve (red curve) at the frequencies of 8GHz, 9GHz, 11GHz, and 12GHz respectively. In other words, two image points cannot be distinguished, indicating that the resolution of square conventional MFELs with $n_0 = 1$ is not enough. Oppositely, the square MFELs whit $n_0 = 2.4$ can resolve the two image points at the frequencies of 8GHz, 9GHz, 11GHz, and 12GHz respectively, as shown in Figs. 4(a)-(d). The relative electric field intensity curve at the imaging plane also reveals that the solid immersion square MFEL with $n_0 = 2.4$ have a super-resolution imaging ability to resolve the two identical point sources in a deep subwavelength distance. Although the WGM effect is stronger and the electric field intensity curve at the imaging position shows obvious resonance fluctuation, it slightly affects the resolution imaging effect of solid immersion square MFEL at the frequency is 11GHz, as shown in Fig. 4(c). Similarly, Figs. 4(a, b, d) show the electric field distribution at the corresponding frequencies of 9 GHz, 11 GHz and 12 GHz, respectively. From the figures, it shows that solid immersion square MFELs with $n_0 = 2.4$ have a very good ability of super-resolution imaging by capturing large wave-number evanescent wave components at the imaging plane to improve the resolution. Consequently, the solid immersion square MFELs with $n_0 = 2.4$ overcome the diffraction limit to achieve super-resolution.

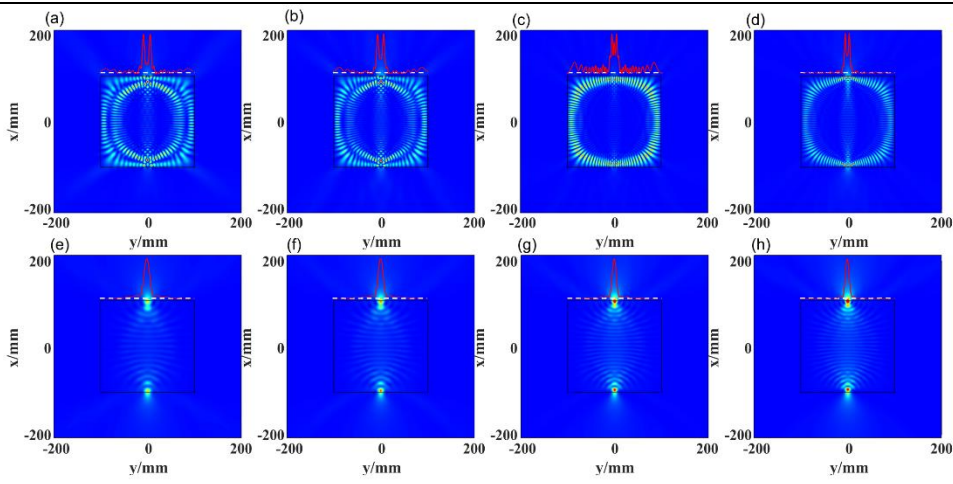

**Fig. 4.** (a)-(d) The imaging performance of solid immersion square MFELs with $n_0 = 2.4$ at the frequencies of 8GHz, 9GHz, 11GHz and 12GHz, respectively. (e)-(g) The imaging performance of conventional square MFELs with $n_0 = 1$ at the frequencies of 8GHz, 9 GHz, 11 GHz and 12GHz, respectively.

Finally, we discuss the further application about solid immersion square MFEL. The solid immersion square MFEL is cascaded to form a super-resolution information transmission channel for potential optical communication, as shown in Fig. 5. Three identical solid immersion square MFELs with $n_0 = 2.4$ are arranged into a row to build up a transmission channel with an air gap of 0.6mm between each solid immersion square MFELs. A pair of point sources are placed at $x = -100$ mm with a distance of $\lambda/3$ to excite TE waves at the frequency of 12GHz. The red curve describes the electric field intensity at different imaging planes with $y = -100 \sim 100$ mm. The size of the air gap in the transmission channel composed of solid immersion square MFELs is determined by referring to the numerical calculated results as shown in Fig. S2 (see Supplement 2 in Supplemental Document). When only one point source is placed in front of the super-resolution information transmission channel, through the observation of the imaging performance of three solid immersion square MFELs, it can be easily found that the information transmission effect and super-resolution performance of the designed super-resolution information transmission channel is relatively good when the air gap spacing is 0.6mm as shown Fig.S2 (see Supplement 2 in Supplemental Document). Generally, when the image point is far away from the source point, the resolution effect will become worse. However, according to the electric field intensity distribution in Fig. 5, it can be seen that even if the distance between the imaging point and source point is increased, the solid immersion square MFELs can still overcome the diffraction limit within a certain range and maintain a good ability of super-resolution imaging. The evanescent wave component with large wave number can be collected at the imaging plane to improve the resolution. Based on the solid immersion square MFEL's super-resolution effect, we design the channel for super-resolution imaging information transmission with larger capacity, which has a good prospect of optical communication applications.

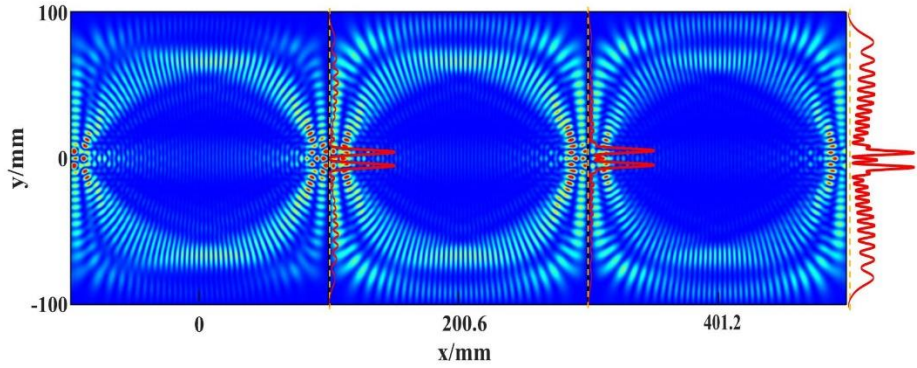

**Fig. 5.** The electric field distribution of super-resolution information channel cascaded by three identical solid immersion MFELs with a 0.6mm air gap at 12 GHz.

### 3. Conclusion

In summary, the circular solid immersion MFEL is transformed into a solid immersion square MFEL using a SC mapping. The solid immersion square MFEL maintains the super-resolution imaging characteristics of the original circular MFEL, and at the same time can overcome the disadvantage of curve surface, which may pave the way to further application such as real-time imaging, vivo imaging and photolithography. Through simulated calculations, the designed square MFEL can achieve super-resolution imaging at most frequencies. Moreover, the cascading solid immersion square MFELs can effectively transmit super-resolution information and have a good prospect of optical communication applications. With the development of 3D printing technology [40, 41], the solid immersion MFEL is expected to be fabricated in high frequency band, such as THz, IR, or even visible frequencies.

### Acknowledgements

National Natural Science Foundation of China (Grant No. 92050102); National Key Research and Development Program of China (Grant No. 2020YFA0710100); National Natural Science Foundation of China (Grant No. 11874311); Fundamental Research Funds for the Central Universities (Grant No. 20720200074).

# Square Maxwell's fish-eye lens for broadband achromatic super-resolution imaging：supplemental document

**Supplement 1: Optimimal design of square Maxwell's fish-eye lens**

In this section, we will propose a new quasi-square Maxwell's fish-eye lens (MFEL) design to solve the problem that the resolution of an original square MFEL is affected by whispering gallery mode (WGM) resonance, which is easily generated due to the four corners. To optimize the design and improve the resolution of solid immersion MFEL, the circular MFEL is transformed into quasi-square MFEL by a method of SC mapping, as shown in Fig. S1(d). The quasi-square MFEL avoids the disadvantage caused by the right angle of a square MFEL, mainly through bending the four angles from straight to curved. Next, we will numerically calculate the imaging performance of the quasi-square MFEL and square MFEL with $n_0=2.4$ by COMSOL Multiphysics, where a point source is placed at $x=-100$ mm and $y=0$ mm as an exciting source. Figs. S1(d)-(f) show the electric field intensity and the corresponding FWHM of the solid immersion quasi-square MFEL at frequencies of 8GHz, 10GHz and 12GHz respectively. To better illustrate the optimized imaging effect, the electric field intensity and the corresponding FWHM of the solid immersion square MFEL at frequencies of 8GHz, 10GHz and 12GHz are also shown in Figs. S1(a)-(c). By comparing the electric field intensity distribution Figs. S1(a)-(c) and (b)-(f), it can be found that the solid immersion quasi-square MFEL significantly avoids distribution of the WGM resonance, and only symmetrical focusing imaging is performed. Consequently, the solid immersion quasi-square MFEL surmounts the disadvantage and achieves better resolution than that of the solid immersion square MFEL. It is proved that the quasi-square MFELs have further optimized the super-resolution effect.

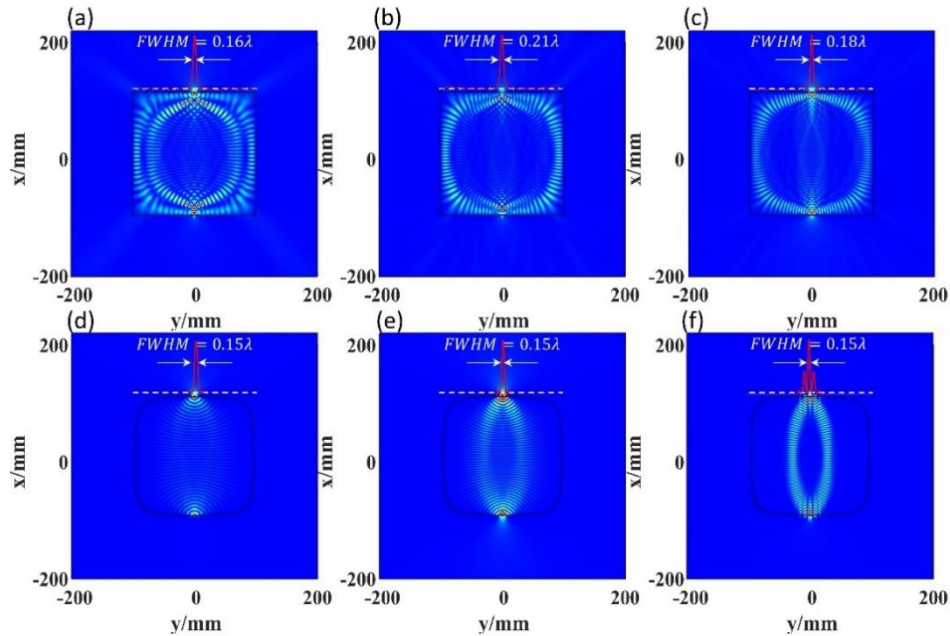

**Fig.** S1. (a)-(f) The electric field intensity distribution and the corresponding FWHM of the solid immersion square MFEL and solid immersion quasi-square MFEL at frequencies of 8GHz, 10GHz and 12GHz, respectively.

**Supplement 2: Numerical analysis of the super-resolution information transmission channel composed of the solid immersion square MFELs**

To design a super-resolution information transmission channel with a better transmission performance, the influence of the air gap between the components (solid immersed square MFELs) of the super-resolution information transmission channel on the transmission channel is discussed and numerical analysis is carried out by using the full-wave simulation method in this section. Three solid immersion square MFELs is arranged in a row to make up a super-resolution information transmission channel, and the spacing $d$ (air slot) between each square MFELs is taken as a variable, as shown in Fig. S2(a). Here the air gap between each solid immersion square MFELs is the same, which is equal to $d$. In the figure, the red dotted line is the observation position of the imaging point ( $y = -100 \sim 100$ mm), which is in the middle of the gap of lens A, B and C respectively. It is the same distance from the imaging boundary of each lenses. The red dots represent a point source which is placed at $x = -100$ mm and $y = 0$ mm to excite the TE cylindrical wave at frequency of 12GHz. According to the super-resolution information transmission channel structure diagram composed of three solid immersion square MFELs as shown in Fig. S2(a), the relative FWHM value is obtained by calculating the electric field intensity distribution along the red dotted line of each square MFELs with the spacing $d$ increasing as shown in Figs. S2(b)-(d). It is clearly observed that the variation curve of FWHM value at each solid immersion square MFELs imaging position concerning air gap spacing $d$ from 0mm to 1.8mm is calculated respectively. Fig. S2(b) indicates the relative value FWHM of solid immersion square MFEL A increases slowly with the increase of air gap $d$, which are basically all below 0.18 $\lambda$. This is because the coupling between solid immersion square MFELs decreases with the increase of air gap $d$, thus, when the air gap $d$ is large, the FWHM of lens A is mainly influenced by total internal reflection (TIR) which is closer to the value of a single solid immersion square MFEL. In Fig. S2(c), it can be seen that the relative value FWHM of the solid immersed square MFEL B reveal a similar "s"-shape change trend when the air gap $d$ increases, that is, the FWHM curve increases firstly and then decline, immediately it rises again, which is affected by the coupling effect between square MFELs and the TIR mechanism. The FWHM's variation trend of solid immersed square MFEL C concerning air gap $d$ is similar to that of lens B.

Through the comprehensive observation of Figs. S2(b)-(d), it can be found that when $d$ is 0 and 0.6mm, the total transmission effect of the super-resolution information transmission channel is relatively good. At the same time, when the air gap is 0.6mm, the channel not only has better information transmission capabilities, but also has super-resolution capabilities. Thus, when the optimal spacing $d$ is 0.6mm, a super-resolution information transmission channel composed of three solid immersion square MFELs is designed. The corresponding FWHM obtained for three solid immersion square MFEL of A, B and C is less than 0.18 $\lambda$, which is less than the diffraction limit. In the main text, Fig. 5 shows the electric field intensity of the designed super-resolution information transmission channel excited by the two point sources. This design has a good optical applications prospect, not only can realize the multi-points source resolution information transmission, but also has the advantage of long-distance transmission.

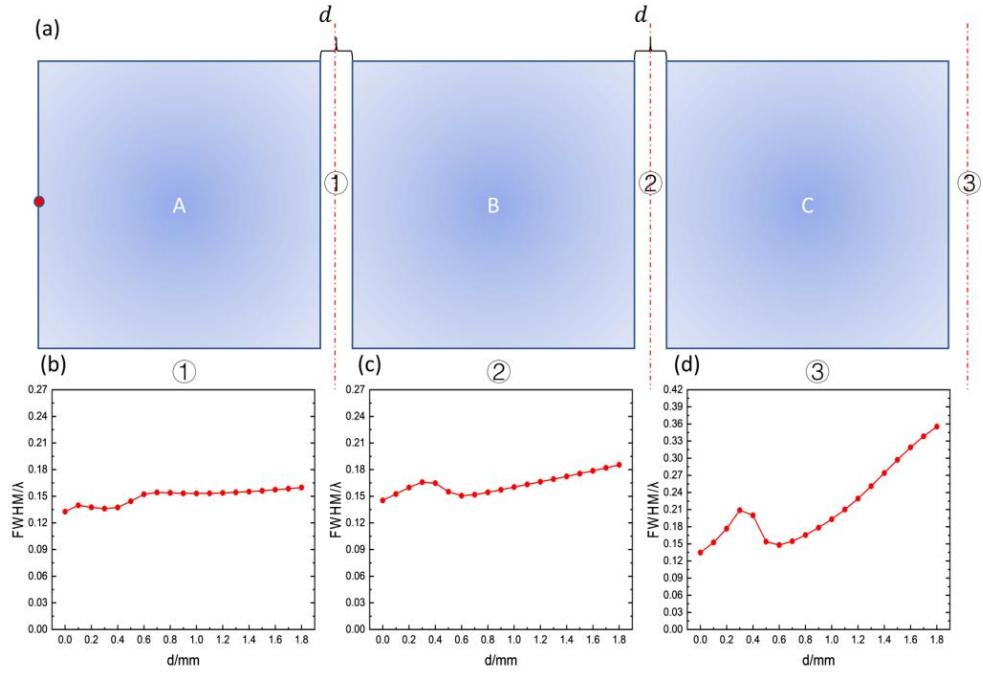

**Fig.** S2. (a) A super-resolution information transmission channel. (b) The curve of FWHM value corresponding to the solid immersion square MFEL A with air spacing $d$. (c) The curve of FWHM value corresponding to the solid immersion square MFEL B concerning air spacing $d$. (d) The curve of FWHM value corresponding to the solid immersion square MFEL C concerning air spacing $d$.